%% file: PVDIS_res_v1.3.tex
\documentclass[aps,prl,showpacs,twocolumn,preprintnumbers,superscriptaddress]{revtex4}
\usepackage{graphicx}% Include figure files
\usepackage{amsfonts, amssymb}
\usepackage{times}
\usepackage{everypage}
\usepackage{gensymb}

\begin{document}

\title{Measurements of Parity-Violating Asymmetries in 
Electron-Deuteron Scattering in the Nucleon Resonance Region}

\input{pvdis_authorlist}

\date{\today}% 

%\end{center}
\begin{abstract}               
We report on parity-violating asymmetries in the nucleon resonance region
measured using inclusive inelastic scattering of 5-6 GeV longitudinally 
polarized electrons off an unpolarized deuterium target. 
These results are the first parity-violating asymmetry data 
in the resonance region beyond the $\Delta(1232)$. They provide a verification
of quark-hadron duality -- the equivalence of the quark-
and hadron-based pictures of the nucleon -- at the (10-15)\% level 
in this electroweak observable, which is dominated by contributions from 
the nucleon electroweak $\gamma Z$ interference structure functions.
In addition, the results 
provide constraints on nucleon resonance models 
%are of particular interest to models 
relevant for calculating background corrections 
%the $\gamma Z$ interference corrections 
to elastic parity-violating electron scattering measurements.
\end{abstract}

 % http://www.aip.org/pacs/

\maketitle

While QCD is the well-established theory of the strong nuclear force,
it remains a challenge to describe the transition from quark and
gluon to hadron degrees of freedom.  
Measurements of the structure functions 
in electron scattering from nuclei, spanning from the low invariant
mass regime ($W < 2$ GeV) of resonance production to the 
deep inelastic scattering (DIS) regime, aim to bridge this transition.
Inclusive measurements from nucleons have
demonstrated a remarkable feature called ``quark-hadron duality'',
first pointed out by Bloom and Gilman~\cite{Bloom:1970xb}, 
in which the low-energy (few GeV) cross sections averaged over
the energy intervals of the resonance structures
resemble those at asymptotically high energies. 
Over the past decade, duality has been verified in 
the unpolarized structure functions $F_2$ and 
$F_L$ at four-momentum-transfer-squared $Q^2$ values 
below 1~(GeV/$c$)$^2$~\cite{Niculescu:2000tk,Liang:2004tj,Psaker:2008ju,Malace:2009dg,Malace:2009kw}, 
the proton spin asymmetry $A_1^p$ down to 
$Q^2=1.6$~(GeV/$c$)$^2$~\cite{Airapetian:2002rw}, 
the spin structure function $g_1$ down to 
$Q^2=1.7$-$1.8$~(GeV/$c$)$^2$~\cite{Bosted:2006gp,Solvignon:2008hk},
the helicity-dependent structure functions $H_{1/2,3/2}$~\cite{Malace:2011ad},
and for charged pion electroproduction in semi-inclusive 
scattering~\cite{Navasardyan:2006gv}.
It was speculated that duality is a universal feature of quark-hadron
transition that should be exhibited not only in electromagnetic interactions,
but also in charged lepton scattering via the weak interaction~\cite{Carlson:1993wy}, and perhaps other processes as well. 
Soon after duality was first observed, attempts were made to understand it from 
the first principles of QCD~\cite{qcd_qhd}, and is even more desired now 
given such solid experimental verification.
For a recent review of both the experimental and theoretical status of duality, 
see Ref.~\cite{Melnitchouk:2005zr}.
Establishing duality, either experimentally or theoretically, also has
practical advantages for the study of nucleon structure. For example, 
the valence quark structure which is typically difficult 
to explore due to the high $Q^2$ required in DIS, may
be studied alternatively by averaging resonance data at lower 
$Q^2$ values~\cite{Malace:2009dg,Malace:2009kw,Malace:2011ad,Arrington:2003nt,12gevA1n_proposal}. 

To study quark-hadron duality in weak interactions, 
it is natural to start with
% the electroweak $\gamma Z$ interference structure functions accessible through 
parity-violating electron scattering (PVES) asymmetries
$A_{PV} = ({\sigma_R} - {\sigma_L})/({\sigma_R} + {\sigma_L})$,
where $\sigma_{R(L)}$ is the cross-section for electrons
polarized parallel (anti-parallel) to their momentum.
The PVES asymmetry on a nucleon 
or nuclear target is dominated by the electroweak 
$\gamma Z$ interference structure functions~\cite{Cahn:1977uu}: 
\begin{eqnarray}
A_{PV} =\left(\frac{G_FQ^2}{4\sqrt{2}\pi \alpha}\right)
  \left(2g_A^e Y_1\frac{F_1^{\gamma Z}}{F_1^\gamma}+{g_V^e}Y_3\frac{F_3^{\gamma Z}}{F_1^\gamma}\right).\label{eq:Apvdis}
\end{eqnarray}
Here $G_F$ is the Fermi weak coupling constant, 
$\alpha$ is the fine structure constant, $Y_1$ and $Y_3$ are 
kinematic factors, %$x$ is the Bjorken scaling variable, 
$g_{V,A}^e$ are the $e-Z^0$ vector and axial couplings, 
and $F_{1,3}^{\gamma,\gamma Z}$ are the electromagnetic and the $\gamma Z$ 
interference structure functions. Note that the $\gamma Z$ functions depend
also on $g_{V,A}^q$, the quark$-Z^0$ vector and axial couplings. 
In the Standard Model, the electron (quark) vector and axial couplings 
are related to the electron's (quark's) quantum numbers and the weak mixing 
angle $\sin^2\theta_W$. 
In practice, the structure functions $F_{1,3}^{\gamma,\gamma Z}$ are
calculated using either parton distribution functions (for deep inelastic
scattering) or nucleon and nuclear models (for elastic scattering or 
nucleon resonances), which provide predictions for asymmetries that
can be compared with the measured values, to either allow extraction of
electroweak parameters such as $\sin^2\theta_W$, or to test models 
used in structure function calculations.
The first PVES experiment~\cite{Prescott1978} 
%performed more than thirty years ago at SLAC \cite{Prescott:1978tm,Prescott:1979dh}, 
provided the first measurement of $\sin^2\theta_W$, and 
established the $SU(2)\times U(1)$ gauge model of Weinberg, Glashow, and 
Salam~\cite{GSW_model} as the correct theory for electroweak interactions.
In the past decade, with the increasing precision accessible to modern  
experiments~\cite{Armstrong:2012bi}, PVES has become a powerful tool to measure
not only $\sin^2\theta_W$, but also $g_{A,V}^{e,q}$ 
% the neutral weak couplings of electrons and quarks 
through DIS measurements~\cite{pvdis_proposal}, 
the nucleon strange form factors via 
elastic scattering~\cite{SAMPLEsff,HAPPEXsff,Aniol:2004hp,G0sff,A4sff}
(for a review see Ref.~\cite{Paschke:2011zz}), 
the weak charge and neutron densities of nuclei 
\cite{Abrahamyan:2012gp,Horowitz:2012tj}, 
and possibly isospin symmetry violation in the 
parton distribution functions (PDFs)~\cite{Londergan:2009kj,Cloet:2012td}.
However, measurements of the PVES asymmetry in the nucleon 
resonance region are scarce. The only existing data are from the 
$G0$ experiment, in which the asymmetry was measured from a proton
target near the $\Delta(1232)$ region with statistical and systematic 
uncertainties of approximately $15\%$ each~\cite{g0_ndelta}.

Measurements of PVES asymmetries in the resonance region 
will also help to test our understanding of the structure of nucleon resonances.
In the resonance region, the PV structure functions 
can be described in terms of longitudinal, transverse, and axial 
PV response functions to specific resonance states, 
together with a non-resonant background. These electroweak 
structure functions can be 
decomposed in terms of their isospin content, providing new and unique
sensitivity to combinations of quark currents weighted 
by their electroweak couplings to the incident electrons~\cite{PV_res}.
%Existing data or calculations for the proton can be extended to 
%the neutron by isospin rotations. 
%%Furthermore, in a simplified picture of the deuteron, the proton
%%and neutron can be considered as nearly free due to the 2.2-MeV
%%binding energy, much smaller than the several-GeV
%%incident electron energy, therefore the asymmetry for the deuteron
%%can be calculated by simple combinations of the proton and the neutron
%%asymmetries. 
%Furthermore, the asymmetry for the deuteron
%can be approximated by simple combinations of the proton and the neutron
%asymmetries due to the its low binding energy.
The asymmetry for the
first nucleon resonance, the $N \rightarrow \Delta(1232)$ transition, was
first calculated by Cahn and Gilman~\cite{Cahn:1977uu}.
Subsequently, more precise calculations in the resonance region have been 
performed~\cite{PV_res}.  
Based on these calculations, the $\Delta(1232)$ asymmetry from the proton 
reported by $G0$ was used to extract the axial form 
factor $G^A_{N \Delta}$~\cite{g0_ndelta}.

In this Letter, we present parity-violating 
asymmetries for scattering longitudinally polarized 
electrons from an unpolarized deuterium target
at four combinations of $Q^2$ 
and invariant mass $W$ spanning the whole nucleon resonance region,
obtained during a recent experiment~\cite{pvdis_proposal} at the 
Thomas Jefferson National Accelerator Facility (JLab). 
These results provide a test of local quark-hadron duality 
in the nucleon electroweak $\gamma Z$ interference 
structure functions and are compared to the theoretical models of 
Matsui, Sato, and Lee~\cite{Matsui:2005ns}, 
Gorchtein, Horowitz, and Ramsey-Musolf~\cite{Gorchtein:2011mz},
and the Adelaide-JLab-Manitoba collaboration~\cite{AJM}.
These results also provide constraints for nucleon resonance models
relevant for calculating background corrections 
%calculating the $\gamma Z$ box diagram corrections 
to elastic PVES~\cite{Gorchtein:2008px,Sibirtsev:2010zg,blunden1and2,Rislow:2010vi,
Gorchtein:2011mz,AJM}.

The experiment was performed in experimental Hall A of JLab.  
A 100-105$\mu$A polarized electron beam 
was incident on a liquid deuterium target and scattered events were detected 
by the Hall A high resolution spectrometer (HRS) pair~\cite{Alcorn:2004sb}
in inclusive mode. 
The main goal of the experiment was
to provide precision PV asymmetries in the DIS region as a test
of the Standard Model~\cite{Beringer:1900zz}
% and to investigate the possibility of extracting 
and to extract the quark weak axial charges $C_{2q}$~\cite{pvdis_proposal}; 
those measurements will be reported in future publications.  
The results reported here come from additional data collected in the
nucleon resonance region during this experiment:
kinematics I-IV were centered
at $W=1.263$, $1.591$, $1.857$, and $1.981$~GeV, respectively.
The $Q^2$ values were just below 1~(GeV/$c$)$^2$ except for kinematics
IV which was at $Q^2=1.472$~(GeV/$c$)$^2$.
The beam energies were 4.867 GeV for kinematics I-III and 6.067 GeV 
for IV.

The polarized electron beam was produced by illuminating a strained 
GaAs photocathode with circularly polarized laser light. The helicity of the
electron beam was selected from a pseudorandom~\cite{HAPPEXsff} 
sequence every 66~ms, and reversed in the middle of this time window, 
forming helicity pairs. The data acquisition was gated by this helicity sequence.
%The helicity sequence controlled the data collection, and periods of beam 
%instability due to helicity reversal were rejected from the data stream. 
To reduce possible systematic errors, a half-wave plate was inserted
intermittently into the path of the polarized laser, which resulted in a 
reversal of the actual beam helicity while keeping the helicity sequence 
unchanged. The expected sign flips in the measured asymmetries between the two 
beam half-wave-plate configurations were observed.
The laser optics of the polarized source were carefully configured to 
minimize changes to the electron-beam parameters under polarization 
reversal~\cite{Sinclair2007}.
A feedback system~\cite{Paschke:2007zz} was used to maintain the 
helicity-correlated intensity asymmetry of the beam below 0.1~parts 
per million (ppm) averaged over the whole experiment. 
The target was a 20-cm-long liquid deuterium cell, with up- and downstream
windows made of 0.10- and 0.13-mm-thick aluminum, respectively.
%
%A luminosity monitor~\cite{HAPPEXsff} located downstream of the target region 
%measured an asymmetry less than 0.1~ppm integrated through the whole experiment.
%
%The helicity-correlated density fluctuation of the target was 
%measured by a luminosity monitor and was found to be below 100 parts per 
%billion (ppb) integrated through the whole experiment.  

In order to count the up-to-600-kHz electron rate and reject the pion 
photo- and electroproduction backgrounds,
a data acquisition (DAQ) and electronic system was specially designed
for this experiment and formed both electron and pion triggers.
%A CO$_2$ gas \v{C}erenkov detector and a double-layered lead-glass 
%shower counter were used to separate electrons from the pion background, 
%and b
The design of the DAQ, along with its particle identification performance and the
deadtime corrections to the measured asymmetries, 
was reported elsewhere~\cite{Subedi:2013jha}.
The overall charged pion $\pi^-$ contamination 
was found to contribute less than $4 \times 10^{-4}$ of the detected
electron rate.
%, with an electron detector 
%efficiency $\geqslant 96\%$ for kinematics I-III and $\geqslant 84\%$ for IV.
Using the measured asymmetries from the pion triggers, the relative uncertainty on 
the measured electron asymmetries $\Delta A/A$ due to the $\pi^-$ background was evaluated to be less than $5\times 10^{-4}$. 
Relative corrections on the asymmetry due to DAQ deadtime were 
$(0.7$-$2.5)\%$ with uncertainties $\Delta A/A<0.5\%$.
%
%The systematic uncertainties from the pion contamination and
%deadtime were negligible compared to the statistical error.
%False asymmetries were found to be consistent with zero from measurements of
%polarized beam scattering off unpolarized $^{12}$C targets.
The standard HRS DAQ~\cite{Alcorn:2004sb} was used at low beam currents to precisely 
determine the kinematics of the experiment. This was realized through
dedicated measurements on a carbon multifoil target which provided data to
determine the transport function of the HRSs. 

The number of scattered particles in each helicity window was normalized
to the integrated charge from the beam current monitors, from which the
raw asymmetries $A_\mathrm{raw}$ were formed. 
%The uncertainty introduced by the beam charge normalization was 
%%estimated
%%using the $\leqslant$3\% 
%%nonlinearity in the beam current monitors~\cite{HAPPEXsff,Aniol:2004hp}
%%multiplied by the $\ll 100$~ppb residual charge asymmetry, and is 
%found to be 
%negligible compared to the statistical uncertainty and was not included
%in the uncertainty evaluation. 
%%
The raw asymmetries were then corrected for helicity-dependent 
fluctuations in the beam parameters, following 
%
%\begin{equation}
$A^\mathrm{bc}_\mathrm{raw}=A_\mathrm{raw}-\sum c_i\Delta x_i$,
%\end{equation}
%
where $\Delta x_i$ are the measured helicity window differences in the beam
position, angle, and energy. 
The values of the correction coefficients $c_i$ could be extracted either from natural movement of the beam, 
or from calibration data collected during the experiment, in which the beam
was modulated several times per hour 
using steering coils and an accelerating cavity. 
The largest of the corrections was approximately $0.4$~ppm, and the difference
between the two methods was used to estimate the systematic uncertainty in the 
beam corrections. 
%These beam corrections and their uncertainties were found to be
%negligible compared to our statistical uncertainties.

The beam-corrected asymmetries $A^\mathrm{bc}_\mathrm{raw}$ were then 
corrected for the beam polarization.
The longitudinal polarization of the electron beam was measured
intermittently during the experiment by a M{\o}ller 
polarimeter~\cite{Alcorn:2004sb}, with a result of $P_b=(90.40\pm 1.54)\%$
for kinematics I-III and $(89.88\pm 1.80)\%$ for IV. In both cases, 
the uncertainty was dominated by the knowledge of the 
M{\o}ller target polarization.
The Compton polarimeter~\cite{Compton_pol} 
measured $(89.45\pm 1.71)\%$ for kinematics IV
%
%monitored the 
%polarization throughout the DIS 
%measurement, but was only available for resonance IV, 
%giving $(89.45\pm 1.71)\%$ 
where the uncertainty
came primarily from the limit in understanding the analyzing power, 
but was not available for kinematics I-III. 
The M{\o}ller and Compton
measurements for kinematics IV were combined to give $(89.65\pm 1.24)\%$. 
The passage of the beam through material before scattering 
causes a small depolarization effect that was corrected. 
This was calculated based on 
Ref.~\cite{Olsen:1959zz} and the beam depolarization was found to be 
less than $6.1\times 10^{-4}$ for all resonance kinematics.

Next, the asymmetries were corrected for various backgrounds. 
The pair-production background, which results from $\pi^0$
decays, was measured at the DIS kinematics of this experiment by reversing 
the polarity of the HRS magnets and was found to contribute less than 
$5\times 10^{-3}$ of the detected rate. Since pion production is smaller
in resonance kinematics than in DIS, and based on the fact that pions
were produced at lower $Q^2$ than electrons of the same momentum and hence
typically have smaller PV asymmetries, the relative uncertainty on the 
measured asymmetries due to this background 
was estimated to be no more than $5\times 10^{-3}$.
Background from the aluminum target windows was estimated using 
Eq.~(\ref{eq:Apvdis}), with structure functions $F_{1,3}^{\gamma Z}$ for aluminum 
constructed from the MSTW DIS PDF~\cite{Martin:2009iq} extrapolated to the 
measured $\langle Q^2\rangle$ and $\langle W\rangle$ values, 
and the latest world fit on the ratio of 
longitudinal to transverse virtual photon electromagnetic absorption
cross sections $R\equiv \sigma_L/\sigma_T$~\cite{Bosted:2007xd}. 
Assuming that the
actual asymmetries differ by no more than 20\% from calculated values
due to resonance structure and nuclear effects, 
the relative correction to the asymmetry is at the 
(1-3)$\times 10^{-4}$ level with an uncertainty 
of $\Delta A/A=0.4\%$ for all kinematics.
Target impurity adds about 0.06\% of relative uncertainty to the measured 
asymmetry due to the presence of a small amount of hydrogen deuteride.
Background from events rescattering off the inner walls of the HRS
was estimated using the probability of such rescattering
%, which was measured
%to be no more than 1\% for the momentum range of $<10\%$ of the HRS
%central momentum, and drops quickly to below 10$^{-3}$ outside this range.
%Since the HRS momentum acceptance is $\pm 4.5\%$, rescattering background
and adds no more than 1\% relative uncertainty to the measured asymmetry.

Corrections from the beam polarization in the direction perpendicular to 
the scattering plane can be described as 
$\delta A=A_n\left[-S_H\sin\theta_{tr}+S_V\cos\theta_{tr}\right]$ where 
$A_n$ is the beam-normal asymmetry,
$S_{V,H,L}$ are respectively the electron polarization components 
in the vertical, % (perpendicular to the nominal scattering plane defined 
%by the electron beam and the central ray of the HRS), 
horizontal %(within the nominal plane but transverse to the beam), 
and 
longitudinal directions,
and $\theta_{tr}$ is the vertical angle of the scattered
electrons.
During the experiment the beam spin components were controlled 
to $\vert S_H/S_L\vert\leqslant 27.4\%$ % use ctan of Wien angle 74.7 deg
and $\vert S_V/S_L\vert\leqslant 2.5\%$ %see Joe's email on 2013/2/27, but could be 2.45%
and the value of $\theta_{tr}$ was 
found to be less than 0.01~rad. Therefore the beam
vertical spin dominates this background: $\delta A\approx A_n S_V\cos\theta_{tr}\leqslant(2.5\%)P_b A_n$ where $P_b=S_L$ is the beam longitudinal polarization 
described earlier. The values of $A_n$ 
were measured at DIS kinematics and 
%should in principle be measured from 
%dedicated measurements with the 
%beam spin aligned vertically to the scattering plane.
%This was performed for the DIS but not for the resonance kinematics.
%However, $A_n$ in the DIS region were found to be 
%no more than 50~ppm and are 
were found to be consistent with 
previous measurements from electron elastic scattering from the proton 
and heavier nuclei~\cite{Abrahamyan:2012cg}.
Based on this 
% which reported that $A_n$ can be approximated 
%as $A_n=(-21.9\sim -32.8)(QA)/Z$~ppm/GeV. For the resonance kinematics reported 
%here this indicates 
it was estimated that, for resonance kinematics, 
$A_n$ varies between $-38$ and $-80$~ppm depending on
the value of $Q^2$, and its amplitude is always smaller 
than that of the corresponding measured electron asymmetry. Therefore the 
uncertainty due
to $A_n$ was estimated to be no more than $2.5\%$ of the measured 
asymmetries. 

Radiative corrections were performed for both internal and external 
bremsstrahlung as well as ionization loss. External radiative corrections were performed based on the 
procedure first described by Mo and Tsai~\cite{Mo:1968cg}. As inputs to the 
radiative corrections, PV asymmetries of elastic scattering from the deuteron were estimated using Ref.~\cite{deuteron-ff}
%Beise:2004py,Abbott:2000ak,Pollock:1990uv} 
and those from quasielastic scattering were
based on Ref.~\cite{Aniol:2004hp}. 
The simulation used to calculate the radiative correction 
also takes into account the effect of HRS acceptance 
and particle identification efficiency variation across the acceptance.

Box-diagram corrections refer to effects that arise 
when the electron simultaneously exchanges two bosons 
($\gamma\gamma$, $\gamma Z$, or $ZZ$ box) with the target, and 
they are dominated by the $\gamma\gamma$ and the $\gamma Z$ box diagrams. 
For PVES asymmetries, the box-diagram effects include  those from the interference between 
$Z$ exchange 
and the $\gamma\gamma$ box, the interference between $\gamma$ exchange and 
the $\gamma Z$ box, and the effect of the $\gamma\gamma$ box on the 
electromagnetic cross sections.
It is expected that there is at least partial cancellation among these
three terms. 
The box-diagram corrections were estimated to be at the (0-1)\% 
level~\cite{PB:priv}, and a 
%for the DIS kinematics of this experiment, and similar sizes were expected
%for the resonances. Therefore a 
$(0.5\pm 0.5)\%$ relative correction was applied to the asymmetries.
%reported here, there exist no reliable calculation for the box diagram 
%and no correction was made.

Results on the physics asymmetry 
$A_{PV}^\mathrm{phys}$ were formed from 
the beam-corrected asymmetry $A^\mathrm{bc}_\mathrm{raw}$ by 
correcting for the beam polarization $P_b$ and backgrounds with asymmetry
$A_i$ and fraction $f_i$, described above,  
%background fractions $b_i$ with asymmetry $A_i$
using the equation
\begin{equation}
A_{PV}^\mathrm{phys} = {{\left(\frac{A^\mathrm{bc}_\mathrm{raw}}{P_b}-
\sum_{i} A_{i}f_{i}\right)}\over{1-\sum_{i} f_{i}}}~.
\end{equation}
%\begin{equation}
% A_{PV} = (\frac{A^\mathrm{bc}_\mathrm{raw}}{P_b} - A_if_i)/(1-f_i)~.
%\end{equation}
When all $f_i$ are small with $A_i$ comparable to or smaller than 
$A^\mathrm{bc}_\mathrm{raw}$, one can define $\bar f_i=f_i(1-\frac{A_i}{A^\mathrm{bc}_\mathrm{raw}}P_b)$
and approximate 
\begin{equation}
 A_{PV}^\mathrm{phys} \approx \frac{A^\mathrm{bc}_\mathrm{raw}}{P_b}\Pi_i\left(1+\bar f_i\right)~,
\end{equation}
{\it i.e.}, all corrections can be treated as multiplicative. 

\begin{table}[!htp]
  \begin{center}
    \begin{tabular}{c|c|c|c|c}
      \hline\hline
  Kinematics  &   I          &    II      &    III     & IV \\\hline
$E_b$ (GeV)   & $4.867$      & $4.867$    & $4.867$    & $6.067$ \\
HRS           & Left         & Left       & Right      & Left \\
$\theta_0$    & $12.9^\circ$  & $12.9^\circ$ & $12.9^\circ$& $15.0^\circ$ \\
$p_0$ (GeV/$c$)& $4.00$      & $3.66$     & $3.10$     & $3.66$ \\
$\langle Q^2\rangle$ [(GeV/$c$)$^2$]
              &  $0.950$     & $0.831$    & $0.757$    & $1.472$  \\
$\langle W\rangle$ (GeV)
              & $1.263$      & $1.591$    & $1.857$    & $1.981$ \\\hline
\multicolumn{5}{c}{Measured asymmetries with beam-related corrections (ppm)}\\\hline
$A^\mathrm{bc}_\mathrm{raw}$ & $-55.11$    & $-63.75$    & $-54.38$  & $-104.04$ \\
 $\pm\Delta A^\mathrm{bc}_\mathrm{raw}$ (stat.)   &$\pm 6.77$   & $\pm 5.91$  & $\pm 4.47$ & $\pm 15.26$ \\ 
 $\pm\Delta A^\mathrm{bc}_\mathrm{raw}$ (syst.)   &$\pm 0.10$   & $\pm 0.15$  & $\pm 0.24$ & $\pm 0.26$ \\
\hline 
%
%\multicolumn{5}{c}{Corrections and systematic uncertainties}\\
%\hline 
%$P_b$          & $90.40\%$    &  $90.40\%$  &  $90.40\%$  & $89.65\%$ \\
%(syst.)        & $\pm 1.54\%$ & $\pm 1.54\%$& $\pm 1.54\%$& $\pm 1.24\%$ \\\hline
%$1+\bar f_\mathrm{depol}$ & $1.00049$ & $1.00028$  & $1.00093$  & $1.00061$ \\
% (syst.)       & $<10^{-4}$   & $<10^{-4}$  & $<10^{-4}$  & $<10^{-4}$ \\\hline
%$1+\bar f_{\mathrm{Al}}$ & $0.9997$    & $0.9998$   & $0.9999$   & $0.9998$ \\
% (syst.)       &$\pm 0.0042$ &$\pm 0.0042$&$\pm 0.0042$& $\pm 0.0042$ \\\hline
%$1+\bar f_\mathrm{dt}$        & $1.0148$    & $1.0247$   & $1.0209$   & $1.0076$  \\
% (syst.)       &$\pm 0.0006$ &$\pm 0.0023$&$\pm 0.0041$&$\pm 0.0004$ \\\hline
%%$f_\mathrm{PID}$(??)   &  \\
%$1+\bar f_\mathrm{rc}$& $1.1095$    &$1.0205$    &  $1.0005$  & $1.0170$  \\
% (syst.)       &$\pm 0.0352$ &$\pm 0.0207$&$\pm 0.0076$& $\pm 0.0112$\\\hline
%$1+\bar f_\mathrm{box}$       
%                & $1.005$     & $1.005$    & $1.005$    & $1.005$ \\
%  (syst.)       & $\pm 0.005$ & $\pm 0.005$&$\pm 0.005$ & $\pm 0.005$ \\\hline 
%$\Delta \bar f_{\pi^-}$      
%                &$\pm 0.018\%$&$\pm 0.046\%$&$\pm 0.019\%$& $0.003\%$ \\
%$\Delta \bar f_\mathrm{pair}$ 
%                & $0.3\%$    & $0.3\%$   & $0.3\%$   & $0.3\%$\\
%$\Delta \bar f_{A_n}$ 
%                & $2.5\%$  & $2.5\%$   & $2.5\%$  & $2.5\%$\\
%$\Delta Q^2$    &$\pm 0.81\%$& $\pm 0.73\%$&$\pm 0.87\%$& $\pm 3.45\%$ \\
%rescattering    &$< 1\%$  &  $< 1\%$  &  $< 1\%$  &  $< 1\%$  \\
%target purity   &$\pm 0.06\%$  &$\pm 0.06\%$  &$\pm 0.06\%$  &$\pm 0.06\%$ \\
\hline
\multicolumn{5}{c}{Physics Asymmetry Results (ppm)}\\
\hline
%with box correction included
%$A_{PV}$ (ppm) & $-70.05$   & $-73.75$   & $-61.49$  & $-118.97$ \\
%(stat.)        & $\pm 8.61$  & $\pm 6.84$ & $\pm 5.05$ & $\pm 17.45$ \\
%(syst.)        & $\pm 2.89$  & $\pm 2.74$ & $\pm 2.00$ & $\pm 5.76$ \\
%(total)        & $\pm 9.08$  & $\pm 7.37$ & $\pm 5.44$ & $\pm 18.38$ \\
$A_{PV}^\mathrm{phys}$& $-68.97$   & $-74.12$   & $-61.80$  & $-119.56$ \\
$\pm\Delta A_{PV}^\mathrm{phys}$(stat.)& $\pm 8.47$  & $\pm 6.87$ & $\pm 5.08$ & $\pm 17.54$ \\
$\pm\Delta A_{PV}^\mathrm{phys}$(syst.)& $\pm 3.30$  & $\pm 2.84$ & $\pm 2.11$ & $\pm 5.62$ \\
$\pm\Delta A_{PV}^\mathrm{phys}$(total)& $\pm 9.09$  & $\pm 7.43$ & $\pm 5.50$ & $\pm 18.42$ \\
%$\Delta A_\mathrm{phys,total}/A_{PV}$ & $13.2\%$ & $10.0\%$ & $8.9\%$ & $15.4\%$\\
\hline
\multicolumn{5}{c}{Calculations (ppm)}\\
\hline
$A_\mathrm{calc}$~\cite{Matsui:2005ns}
               & $-89.10$    & N/A         & N/A         & N/A        \\\hline
$A_\mathrm{calc}$
               & $-88.94$    & $-70.29$    & $-65.09$    & $-124.74$      \\
$\pm \Delta A_\mathrm{calc}$~\cite{Gorchtein:2011mz}
             &$^{+9.98}_{-8.76}$&$^{+14.81}_{-11.09}$&$^{+11.85}_{-10.95}$& $^{+20.12}_{-19.49}$ \\\hline
$A_\mathrm{calc}$       & $-88.22$& $-69.63$ & $-65.23$& $-124.75$\\
$\pm \Delta A_\mathrm{calc}$~\cite{AJM} 
             &$^{+8.10}_{-8.31}$&$^{+7.05}_{-7.19}$&$^{+5.19}_{-5.34}$&$^{+9.11}_{-9.49}$\\\hline
%$A_\mathrm{DIS}^\mathrm{MSTW}$
%               & $-74.97$    & $-66.75$   & $-61.91$   & $-120.68$ \\\hline
$A_\mathrm{calc}^\mathrm{DIS,CJ}$
               & $-75.63$    & $-66.72$   & $-61.59$   & $-119.13$ \\
%                $-68.97$   & $-74.12$   & $-61.80$  & $-119.56$ \\
%                $\pm 9.09$  & $\pm 7.43$ & $\pm 5.50$ & $\pm 18.42$ \\
%dA(data)/A(DIS) 12.1%    11.13%    8.9%
%$\pm\Delta A_\mathrm{dis}$~(ppm)
%               & $<1$    & $<1$   & $<1$   & $<1$ \\
\hline\hline
      \end{tabular}
  \end{center}
  \caption{Asymmetry results on parity-violating $\vec e-^2$H 
scattering in the nucleon resonance region.  
The kinematics shown include the beam energy $E_b$, with which HRS was 
used (Left or Right), central 
angle and momentum settings of the HRS $\theta_0, p_0$, 
and the actual kinematics averaged from the data $\langle Q^2\rangle$ and 
$\langle W\rangle$. 
The beam-corrected asymmetries $A^\mathrm{bc}_\mathrm{raw}$ are shown
along with their statistical precision and systematic uncertainties due to
beam-related corrections.
%, werecorrected for the effects from the beam polarization $P_b$ 
%and others including:
%the beam depolarization $\bar f_\mathrm{depol}$, 
%the target aluminum endcap $\bar f_{\mathrm{Al}}$,
%the DAQ deadtime $\bar f_\mathrm{dt}$~\cite{Subedi:2013jha},
%%the particle identification $\bar f_\mathrm{PID}$ (???),  
%the radiative correction $\bar f_\mathrm{rc}$, 
%and the box diagram correction $\bar f_\mathrm{box}$.
%Other uncertainties that affect the asymmetries include: 
%the charged pion and the pair production background $\bar f_{\pi^-}$ 
%and $\bar f_\mathrm{pair}$, 
%the beam normal asymmetry $\bar f_{A_n}$, target purity, rescattering background, 
%and the uncertainty in the determination of $Q^2$.
Final results on the physics asymmetries $A_{PV}^\mathrm{phys}$ are 
compared with calculations
from three resonance models~\cite{Matsui:2005ns,Gorchtein:2011mz,AJM} 
as well as %two 
DIS estimations using %MSTW~\cite{Martin:2009iq} and $A_\mathrm{DIS}^\mathrm{MSTW}$ and 
CJ~\cite{Owens:2012bv} PDF fits $A_\mathrm{calc}^\mathrm{DIS,CJ}$.
}
\label{tab:Aresults}
\end{table}

Table~\ref{tab:Aresults} shows all kinematics, the beam-corrected asymmetries
$A^\mathrm{bc}_\mathrm{raw}$, and the final
asymmetry results $A_{PV}^\mathrm{phys}$ 
compared to calculations from Matsui, Sato, and Lee~\cite{Matsui:2005ns} 
[for $\Delta(1232)$ only]; 
Gorchtein, Horowitz, and Ramsey-Musolf~\cite{Gorchtein:2011mz}; 
and the Adelaide-JLab-Manitoba model~\cite{AJM}.
In addition, the structure functions $F_{1,3}^{\gamma(Z)}$ in Eq.~(\ref{eq:Apvdis}) 
can be estimated using PDF fits obtained from DIS data, extrapolated to the 
resonance region, along with 
the quark-$Z^0$ vector and axial couplings $g_{V,A}^q$ 
based on Standard Model values~\cite{Beringer:1900zz}.
%of the quark weak vector and axial charges $C_{1q}=2g_A^e g_V^q$ and 
%$C_{2q}=2g_V^e g_A^q$, 
This approach provides
DIS estimations $A_\mathrm{calc}^\mathrm{DIS}$ that can be compared
to the measured asymmetries to test quark-hadron duality. 
% that were evaluated from the parton distribution functions and the 
%quark-$Z^0$ vector and axial couplings $g_{V,A}^q$. 
For these DIS estimations, 
electroweak radiative corrections were applied to $g_{V,A}^q$ directly, and three
PDF fits -- MSTW~\cite{Martin:2009iq}, CTEQ-Jefferson Lab (CJ)~\cite{Owens:2012bv} and CT10~\cite{Lai:2010vv} 
-- extrapolated to the 
measured $\langle Q^2\rangle$ and $\langle W\rangle$ values were used along 
with world data on $R$~\cite{Bosted:2007xd}.
The uncertainty from each PDF fit was below a fraction of a ppm and 
the differences among all three fits were below 1.5~ppm 
for all kinematics. From Table~\ref{tab:Aresults} one can see that the 
final asymmetry results agree very well with the DIS calculations, 
indicating that for the $Q^2$ range covered by these measurements, 
%for $Q^2$ values below 1~(GeV/$c$)$^2$, 
duality holds throughout the whole resonance region at the 
(10-15)\% level.

\begin{figure}
%color version
 \includegraphics[width=0.5\textwidth]{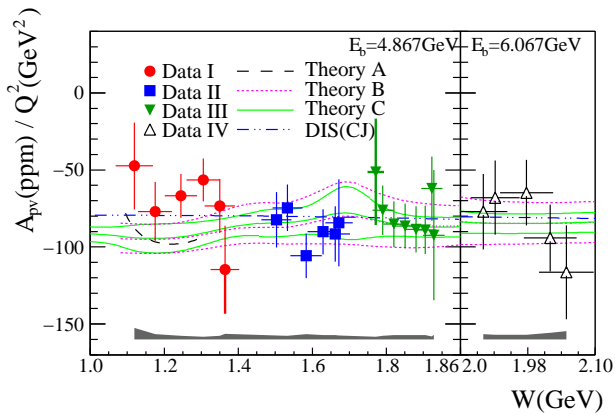}
%black&white version
% \includegraphics[width=0.5\textwidth]{newA_group_W_theory_cj_Wspan_edit.eps}
 \caption{({\it Color online}) 
$W$-dependence of the parity-violating asymmetries in 
$\vec e-^2$H scattering extracted from this experiment. The
physics asymmetry results $A_{PV}^\mathrm{phys}$ for the four kinematics 
I, II, III and IV (solid circles, solid squares, solid triangles, and
open triangles, respectively), in ppm, 
are scaled by $1/Q^2$ and compared with calculations from 
Ref.~\cite{Matsui:2005ns} (theory $A$, dashed lines), 
Ref.~\cite{Gorchtein:2011mz} (theory $B$, dotted lines), 
Ref.~\cite{AJM} (theory $C$, solid lines) and the 
DIS estimation (dash-double-dotted lines) using Eq.~(\ref{eq:Apvdis}) with 
the extrapolated CJ PDF~\cite{Owens:2012bv}. 
The vertical error bars for the data are statistical uncertainties, while
the horizontal error bars indicate the root-mean-square values of the
$W$ coverage of each bin. The experimental
systematic uncertainties are shown as the shaded bands at the bottom. 
For each of the four
kinematics, calculations were
performed at the fixed $E_b$ and $Q^2$ values of Table~\ref{tab:Aresults}
and with a variation in $W$ to match the coverage of the data.
Theories $B$ and $C$ each have three curves showing the central values
and the upper and lower bounds of the calculation. Uncertainties
of the DIS calculation were below 1~ppm and are not visible.
}\label{fig:Apv_gr}
\end{figure}

In addition to the results in Table~\ref{tab:Aresults}, 
asymmetry results
with smaller bins in $W$ are also available due to the
detector segmentation and trigger electronics adopted in this
experiment~\cite{Subedi:2013jha}:  
for each kinematics, six (eight) ``group'' 
triggers were formed first from different segments of the detectors for the
Left (Right) HRS, and 
a logical OR of all group triggers was formed to give
a global trigger. While asymmetry results from the global trigger, 
shown in Table~\ref{tab:Aresults}, provided higher statistical precision, 
asymmetries extracted from group triggers  
allowed study of the detailed $W$-dependence of the asymmetry within each 
kinematic setting, with little variation in $Q^2$.
Figure~\ref{fig:Apv_gr} shows the $W$-dependence of asymmetry results  
$A_{PV}^\mathrm{phys}$, scaled by $1/Q^2$, extracted from group 
triggers. 
%The number of bins for each kinematics
%depends on the detector segmentation at which the data were taken. 
%(6 bins for RES I, II and IV, and 8 bins for RES III). 
The data between
adjacent bins within each kinematics 
typically have a (20-30)\% overlap in event samples and are
thus correlated, while the lowest and the highest bins of each kinematics have
larger overlaps with %, and are closer in $W$ to, 
their adjacent bins. 

% compared with calculations from 
%the three resonance models~\cite{Matsui:2005ns,Gorchtein:2011mz,AJM} 
%as well as the DIS estimation. 
One can see from Fig.~\ref{fig:Apv_gr}
that the measured asymmetries at all kinematics 
are consistent with the three resonance
models, and again agree very well with the DIS estimation. 
No significant resonance structure is observed in the $W$-dependence
of the asymmetries.

In summary, we report here results on the parity-violating asymmetries in the 
nucleon resonance region, including the first PV asymmetry data 
beyond the $\Delta(1232)$ resonance. 
These results provide important constraints to
nucleon resonance models relevant for calculating background corrections 
% $\gamma Z$ box-diagram corrections 
to elastic parity-violating electron scattering measurements.
The agreement with DIS-based calculations indicates
that quark-hadron duality holds for PVES asymmetries on the deuteron at the 
(10-15)\% level throughout the resonance region, 
for $Q^2$ values just below 1~(GeV/$c$)$^2$. 
These results are comparable to 
the unpolarized electromagnetic structure
function data which verified duality at the (5-10)\% level for the 
proton and (15-20)\% for the neutron at similar $Q^2$ values, 
although the unpolarized 
measurements provided better resolution in $W$ and covered a broader
kinematic range~\cite{Malace:2009dg,Malace:2009kw,Malace:2011ad}.
%although for $\Delta(1232)$ the results at kinematics I 
%agrees better with the DIS calculation than the resonance 
%models, although it could also be due to statistical fluctuations.  
%Future measurements, if planned, may help verify 
%quark-hadron duality in the $\gamma Z$ interference structure functions to higher
%precision, and establish further duality to be a universal property of the nucle%on
%structure exhibited by both electromagnetic and weak interactions.
We have therefore provided the first experimental support for the
hypothesis that quark-hadron duality is a universal property of nucleons
in both their weak and their electromagnetic interactions.

\begin{acknowledgments}

The authors would like to thank the personnel of Jefferson Lab for 
their efforts which resulted in the successful completion of the experiment,
and T.-S.~H.~Lee, T. Sato, M.~Gorshteyn, N.~Hall, W.~Melnitchouk and 
their collaborators for carrying out the nucleon resonance calculations.
X.~Zheng would like to thank the Medium Energy Physics Group at the 
Argonne National Lab for supporting her during the initial work of this 
experiment. 
This work was supported in part by the Jeffress Memorial Trust under 
Award No. J-836, the U.S. National Science Foundation under Award No. 0653347, 
and the U.S. Department of Energy under Award No. DE-SC0003885 
and DE-AC02-06CH11357.
{\bf Notice:} Authored by Jefferson Science Associates, LLC under U.S. DOE
Contract No. DE-AC05-06OR23177. The U.S. Government retains a non-exclusive, 
paid-up, irrevocable, world-wide license to publish or reproduce this 
manuscript for U.S. Government purposes.

\end{acknowledgments}

\input{PVDIS_res_v1.3.bib}

\end{document}

%% file: pvdis_authorlist.tex
%
% copied from HAPPEX-III PRL draft
%
% YOU NEED THIS (and not ``superscriptaddress''):
% \documentclass[twocolumn,showpacs,preprintnumbers,amsmath,amssymb]{revtex4}
%
% Within REVTEX4, to get superscripts from multiple authors tied to one 
% institution, while typing the institution name only once, is a bit awkward.  
% We use the ``edef'' macro to define successive superscript numbers in the
% same order as the ``affiliations'' are listed. These superscripts 
% (e.g. \calstate) are then connected to authors.  To add another insitution, 
% put the appropriate ``edef'' line and ``affiliation'' line.
%
\collaboration{The Jefferson Lab Hall A Collaboration}
\noaffiliation
\author{D.~Wang}
\affiliation{University of Virginia, Charlottesville, Virginia 22904, USA}

\author{K.~Pan}
\affiliation{Massachusetts Institute of Technology, Cambridge, Massachusetts 02139, USA}

\author{R.~Subedi} %\thanks{now at George Washington University, Washington, D.C. 20052, USA}
\thanks{now at Richland College, 
Dallas County Community College District, 
%12800 Abrams Rd  
Dallas, Texas 75243, USA.}
\affiliation{University of Virginia, Charlottesville, Virginia 22904, USA}

\author{ X.~Deng}
\affiliation{University of Virginia, Charlottesville, Virginia 22904, USA}

%--------------------------------------------------

\author{Z.~Ahmed}
\affiliation{Syracuse University, Syracuse, New York 13244, USA} 

\author{ K.~Allada}
\affiliation{University of Kentucky, Lexington, Kentucky 40506, USA}

\author{K.~A.~Aniol}
\affiliation{\mbox{California State University, Los Angeles}, Los Angeles, California 90032, USA }

\author{D.~S.~Armstrong}
\affiliation{College of William and Mary, Williamsburg, Virginia 23187, USA}

\author{J.~Arrington}
\affiliation{Physics Division, Argonne National Laboratory, Argonne, Illinois 60439, USA}

\author{ V.~Bellini}
\affiliation{Istituto Nazionale di Fisica Nucleare, Dipartimento di Fisica dell'Universit\`a~di Catania, I-95123 Catania, Italy}

\author{R.~Beminiwattha}
\affiliation{Ohio University, Athens, Ohio 45701, USA}

\author{J.~Benesch}
\affiliation{Thomas Jefferson National Accelerator Facility, Newport News, Virginia 23606, USA} 

\author{F.~Benmokhtar}
\affiliation{Carnegie Mellon University, Pittsburgh, Pennsylvania 15213, USA}

\author{A.~Camsonne}
\affiliation{Thomas Jefferson National Accelerator Facility, Newport News, Virginia 23606, USA} 

\author{M.~Canan}
\affiliation{Old Dominion University, Norfolk, Virginia 23529, USA}

\author{G.~D.~Cates}
\affiliation{University of Virginia, Charlottesville, Virginia 22904, USA}
 
\author{J.-P.~Chen} 
\affiliation{Thomas Jefferson National Accelerator Facility, Newport News, Virginia 23606, USA} 

\author{E.~Chudakov} 
\affiliation{Thomas Jefferson National Accelerator Facility, Newport News, Virginia 23606, USA} 

\author{ E.~Cisbani}
\affiliation{INFN, Sezione di Roma, gruppo Sanit\`a and Istituto Superiore di Sanit\`a, I-00161 Rome, Italy}

\author{M.~M.~Dalton}
\affiliation{University of Virginia, Charlottesville, Virginia 22904, USA}

\author{C.~W.~de~Jager} 
\affiliation{Thomas Jefferson National Accelerator Facility, Newport News, Virginia 23606, USA} 
\affiliation{University of Virginia, Charlottesville, Virginia 22904, USA}

\author{ R.~De Leo}
\affiliation{Universit\`a di Bari, I-70126 Bari, Italy}

\author{W.~Deconinck}
\affiliation{College of William and Mary, Williamsburg, Virginia 23187, USA}

\author{A.~Deur}
\affiliation{Thomas Jefferson National Accelerator Facility, Newport News, Virginia 23606, USA} 

\author{C.~Dutta}
\affiliation{University of Kentucky, Lexington, Kentucky 40506, USA}

\author{L.~El~Fassi}
\affiliation{Rutgers, The State University of New Jersey, Newark, New Jersey 07102, USA}

\author{D.~Flay}
\affiliation{Temple University, Philadelphia, Pennsylvania 19122, USA} 

\author{G.~B.~Franklin}
\affiliation{Carnegie Mellon University, Pittsburgh, Pennsylvania 15213, USA}

\author{M.~Friend}
\affiliation{Carnegie Mellon University, Pittsburgh, Pennsylvania 15213, USA}

\author{S.~Frullani}
\affiliation{INFN, Sezione di Roma, gruppo Sanit\`a and Istituto Superiore di Sanit\`a, I-00161 Rome, Italy}

\author{F.~Garibaldi} 
\affiliation{INFN, Sezione di Roma, gruppo Sanit\`a and Istituto Superiore di Sanit\`a, I-00161 Rome, Italy}

\author{A.~Giusa}
\affiliation{Istituto Nazionale di Fisica Nucleare, Dipartimento di Fisica dell'Universit\`a~di Catania, I-95123 Catania, Italy}

\author{A.~Glamazdin} 
\affiliation{Kharkov Institute of Physics and Technology, Kharkov 61108, Ukraine} 

\author{S.~Golge}
\affiliation{Old Dominion University, Norfolk, Virginia 23529, USA}

\author{ K.~Grimm}
\affiliation{Louisiana Technical University, Ruston, Louisiana 71272, USA}

\author{K.~Hafidi}
\affiliation{Physics Division, Argonne National Laboratory, Argonne, Illinois 60439, USA}

\author{O.~Hansen} 
\affiliation{Thomas Jefferson National Accelerator Facility, Newport News, Virginia 23606, USA} 

\author{D.~W.~Higinbotham} 
\affiliation{Thomas Jefferson National Accelerator Facility, Newport News, Virginia 23606, USA} 

\author{R.~Holmes} 
\affiliation{Syracuse University, Syracuse, New York 13244, USA} 

\author{T.~Holmstrom} 
\affiliation{Longwood University, Farmville, Virginia 23909, USA}

\author{R.~J.~Holt}
\affiliation{Physics Division, Argonne National Laboratory, Argonne, Illinois 60439, USA}

\author{ J.~Huang}
\affiliation{Massachusetts Institute of Technology, Cambridge, Massachusetts 02139, USA}

\author{C.~E.~Hyde}
\affiliation{Old Dominion University, Norfolk, Virginia 23529, USA}
\affiliation{Clermont Universit\'e, Universit\'e Blaise Pascal, CNRS/IN2P3,
Laboratoire de Physique Corpusculaire, FR-63000 Clermont-Ferrand, France}

\author{C.~M.~Jen}
\affiliation{Syracuse University, Syracuse, New York 13244, USA} 

\author{D.~Jones}
\affiliation{University of Virginia, Charlottesville, Virginia 22904, USA}

\author{ H.~Kang}
\affiliation{Seoul National University, Seoul 151-742, South Korea}

\author{P.~King}
\affiliation{Ohio University, Athens, Ohio 45701, USA}

\author{S.~Kowalski}
\affiliation{Massachusetts Institute of Technology, Cambridge, Massachusetts 02139, USA} 

\author{K.~S.~Kumar}
\affiliation{University of Massachusetts Amherst, Amherst, Massachusetts 01003, USA}
 
\author{J.~H.~Lee}
\affiliation{College of William and Mary, Williamsburg, Virginia 23187, USA}
\affiliation{Ohio University, Athens, Ohio 45701, USA}

\author{J.~J.~LeRose} 
\affiliation{Thomas Jefferson National Accelerator Facility, Newport News, Virginia 23606, USA} 

\author{N.~Liyanage}
\affiliation{University of Virginia, Charlottesville, Virginia 22904, USA}
 
\author{E.~Long}
\affiliation{Kent State University, Kent, Ohio 44242, USA} 

\author{D.~McNulty}
\thanks{now at Idaho State University, Pocatello, Idaho 83201, USA.}
\affiliation{University of Massachusetts Amherst, Amherst, Massachusetts 01003, USA}

\author{D.~J.~Margaziotis}
\affiliation{\mbox{California State University, Los Angeles}, Los Angeles, California 90032, USA }

%7 on HAPPEX, 0 on PVDIS... ...
\author{F.~Meddi}
\affiliation{INFN, Sezione di Roma and Sapienza - Universit\`a di Roma, I-00161 Rome, Italy}

\author{D.~G.~Meekins} 
\affiliation{Thomas Jefferson National Accelerator Facility, Newport News, Virginia 23606, USA} 

\author{L.~Mercado}
\affiliation{University of Massachusetts Amherst, Amherst, Massachusetts 01003, USA}

\author{Z.-E.~Meziani} 
\affiliation{Temple University, Philadelphia, Pennsylvania 19122, USA} 

\author{R.~Michaels} 
\affiliation{Thomas Jefferson National Accelerator Facility, Newport News, Virginia 23606, USA} 

\author{M.~Mihovilovic}
\affiliation{Institut Jo\v zef Stefan, 3000 SI-1001 Ljubljana, Slovenia}

\author{N.~Muangma}
\affiliation{Massachusetts Institute of Technology, Cambridge, Massachusetts 02139, USA} 

\author{K.~E.~Myers}
\thanks{now at Rutgers, The State University of New Jersey, Newark, New Jersey 07102, USA.}
\affiliation{George Washington University, Washington, D.C. 20052, USA}

\author{S.~Nanda}
\affiliation{Thomas Jefferson National Accelerator Facility, Newport News, Virginia 23606, USA} 

\author{ A.~Narayan}
\affiliation{Mississippi State University, Starkeville, Mississippi 39762, USA}

\author{V.~Nelyubin}
\affiliation{University of Virginia, Charlottesville, Virginia 22904, USA}
%\affiliation{St.Petersburg Nuclear Physics Institute of Russian Academy of Science, Gatchina, 188350, Russia}

\author{ Nuruzzaman}
\affiliation{Mississippi State University, Starkeville, Mississippi 39762, USA}

\author{ Y.~Oh}
\affiliation{Seoul National University, Seoul 151-742, South Korea}

\author{D.~Parno}
\affiliation{Carnegie Mellon University, Pittsburgh, Pennsylvania 15213, USA}

\author{K.~D.~Paschke}
\affiliation{University of Virginia, Charlottesville, Virginia 22904, USA}

\author{ S.~K.~Phillips}
\affiliation{University of New Hampshire, Durham, New Hampshire 03824, USA}

\author{ X.~Qian}
\affiliation{Duke University, Durham, North Carolina 27708, USA}

\author{ Y.~Qiang}
\affiliation{Duke University, Durham, North Carolina 27708, USA}

\author{B.~Quinn}
\affiliation{Carnegie Mellon University, Pittsburgh, Pennsylvania 15213, USA}

\author{A.~Rakhman}
\affiliation{Syracuse University, Syracuse, New York 13244, USA} 

\author{P.~E.~Reimer}
\affiliation{Physics Division, Argonne National Laboratory, Argonne, Illinois 60439, USA}

\author{ K.~Rider}
\affiliation{Longwood University, Farmville, Virginia 23909, USA}

\author{S.~Riordan}
\affiliation{University of Virginia, Charlottesville, Virginia 22904, USA}
 
\author{J.~Roche} 
\affiliation{Ohio University, Athens, Ohio 45701, USA}

\author{ J.~Rubin}
\affiliation{Physics Division, Argonne National Laboratory, Argonne, Illinois 60439, USA}

\author{ G.~Russo}
\affiliation{Istituto Nazionale di Fisica Nucleare, Dipartimento di Fisica dell'Universit\`a~di Catania, I-95123 Catania, Italy}

\author{K.~Saenboonruang}
\thanks{now at Kasetsart University, %50 Ngam Wong Wan Rd,
%Lat Yao, Chatuchak, 
Bangkok 10900, Thailand}
\affiliation{University of Virginia, Charlottesville, Virginia 22904, USA}

\author{A.~Saha} \thanks{Deceased.}
\affiliation{Thomas Jefferson National Accelerator Facility, Newport News, Virginia 23606, USA} 

\author{B.~Sawatzky}
\affiliation{Thomas Jefferson National Accelerator Facility, Newport News, Virginia 23606, USA} 

\author{A.~Shahinyan} 
\affiliation{Yerevan Physics Institute, Yerevan 0036, Armenia}

\author{R.~Silwal}
\affiliation{University of Virginia, Charlottesville, Virginia 22904, USA}

\author{ S.~Sirca}
\affiliation{Institut Jo\v zef Stefan, 3000 SI-1001 Ljubljana, Slovenia}

\author{P.~A.~Souder}
\affiliation{Syracuse University, Syracuse, New York 13244, USA} 

%6 on HAPPEX, 0 on PVDIS
\author{R.~Suleiman} 
\affiliation{Thomas Jefferson National Accelerator Facility, Newport News, Virginia 23606, USA} 

\author{V.~Sulkosky}
\affiliation{Massachusetts Institute of Technology, Cambridge, Massachusetts 02139, USA} 

%0 on HAPPEX, 7 on PVDIS
\author{ C.~M.~Sutera}
\affiliation{Istituto Nazionale di Fisica Nucleare, Dipartimento di Fisica dell'Universit\`a~di Catania, I-95123 Catania, Italy}

%0 on HAPPEX, 7 on PVDIS
\author{W.~A.~Tobias}
\affiliation{University of Virginia, Charlottesville, Virginia 22904, USA}

%7 on HAPPEX, 0 on PVDIS... ...
\author{G.~M.~Urciuoli}
\affiliation{INFN, Sezione di Roma and Sapienza - Universit\`a di Roma, I-00161 Rome, Italy}

\author{B.~Waidyawansa}
\affiliation{Ohio University, Athens, Ohio 45701, USA}

\author{B.~Wojtsekhowski}
\affiliation{Thomas Jefferson National Accelerator Facility, Newport News, Virginia 23606, USA} 

\author{ L.~Ye}
\affiliation{China Institute of Atomic Energy, Beijing, 102413, People's Republic of China}

\author{ B.~Zhao}
\affiliation{College of William and Mary, Williamsburg, Virginia 23187, USA}

\author{X.~Zheng}\thanks{author to whom correspondence should be addressed}
\email{xiaochao@jlab.org}
\affiliation{University of Virginia, Charlottesville, Virginia 22904, USA}

%% file: PVDIS_res_v1.3.bib.tex
%\vskip .1truein